\documentclass [aps,prl,showpacs,twocolumn] {revtex4}
\usepackage{graphicx}
\begin{document}

\title{\bf Fluctuational transitions through a fractal basin boundary}

\author{A.~N.~Silchenko, S.~Beri, D.~G.~Luchinsky and P.~V.~E.~McClintock}

\affiliation{Department of Physics, Lancaster University,
Lancaster LA1 4YB, UK}

\date{\today}

\begin{abstract}

Fluctuational transitions between two co-existing chaotic
attractors, separated by a fractal basin boundary, are studied in
a discrete dynamical system. It is shown that the
mechanism for such transitions is determined by a
hierarchy of homoclinic points. The most probable escape
path from the chaotic attractor to the fractal boundary is found
using both statistical analyses of fluctuational trajectories and
the Hamiltonian theory of fluctuations.

\end{abstract}


\maketitle

The mechanism of fluctuational escape from a chaotic attractor
(CA) through a fractal basin boundary (FBB) represents one of the
most challenging unsolved problems in fluctuation theory
\cite{Kautz:87,Beale:89,Grassberger:89,Hamm:91}. The unpredictable
and highly complex stochastic behavior of such systems arises in
part from the presence of limit sets of complex geometrical
structure, and in part from the fractality of the basin
boundary~\cite{Guken,Ott}. For this reason, the central question
-- whether or not there exists a generic mechanism for
fluctuational transitions through the FBB -- has remained
unanswered. More specifically, it has remained unclear: (i) if
boundary conditions can be found both on the CA and on the FBB;
(ii) if there exits a unique escape path from the CA to the FBB;
(iii) whether this path can be determined using the Hamiltonian
theory of fluctuations; (iv) if there is any deterministic
structure involved in the transition through the FBB itself; and
(v) what effect is exerted by the noise intensity. If transitions
across FBBs are characterised by general features, a knowledge of
them could considerably simplify analyses of both stability and
control for chaotic dynamical systems, which are problems of broad
interdisciplinary interest~\cite{Fradkov:98,Boccaletti:00}.

A promising approach to the solution of this problem is based on
the analysis of fluctuations in the limit of very small noise
intensity. In this limit, a stochastic dynamical system fluctuates
to remote states along certain most probable deterministic
paths~\cite{Onsager:53,Dykman:92a,Luchinsky:97}, corresponding to
rays in the WKB-like asymptotic solution of the Fokker-Planck
equation~\cite{Freidlin:84}. The possibility of extending such an
approach to chaotic systems, both continuous and discrete, was
established
earlier~\cite{Kautz:87,Beale:89,Grassberger:89,Hamm:91}. It was
shown also that the presence of homoclinic tangencies, causing the
fractalization of the basins, causes a decrease in the activation
energy~\cite{Soskin:01}.

In this Letter we show that a generic mechanism of fluctuational
transition between co-existing CAs separated by an FBB does exist,
that it is determined by a hierarchy of homoclinic original
saddles forming the homoclinic structure, and that
there is a unique most probable escape path (MPEP) from the CA
that approaches an accessible orbit on the fractal boundary.

\begin{figure}
\includegraphics[width=7.2cm,height=6.2cm]{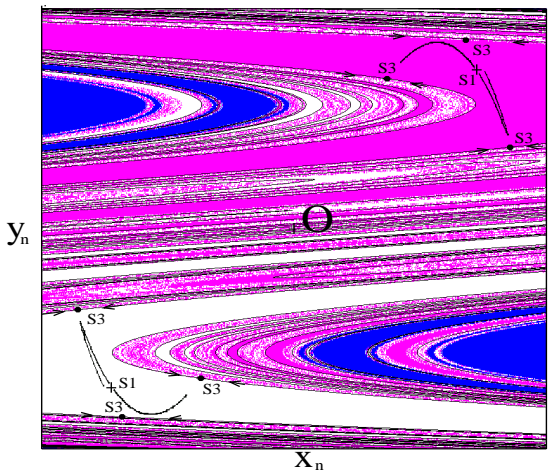}
\caption{\label{fig1} The co-existing chaotic attractors (filled
black regions) and their basins of attraction represented by grey
and white respectively. The accessible boundary saddle points of
period 3 are shown by the small filled black circles labelled S3.
Their stable manifolds are shown as full black lines. The saddle
points of period 1 are shown by the crosses labelled S1. The
saddle point at the origin is labelled O.}
\end{figure}

To demonstrate the existence of this escape mechanism, we take as
an example the two-dimensional map introduced by
Holmes~\cite{Holmes:79}. The properties of this map, including the
structures both of its CA and of its locally disconnected FBB, are
generic for a wide class of maps and flow
systems~\cite{Cartwright:51,Yorke:85a}. It is this fact, taken
with the results of our investigations of escape in other systems,
that allows us to conclude that the escape mechanism described
below is indeed a typical one. The Holmes map is
\begin{eqnarray}
\label{hmap} x_{n+1}&=& y_{n}\\ \nonumber y_{n+1}&=& -b\, x_{n} +
d\, y_{n}-y_{n}^{3}+\xi_n,
\end{eqnarray}
where $\xi_n$ is zero-mean, white, Gaussian noise of variance $D$.
Due to symmetry, the noise-free system (\ref{hmap}) with $b=0.2$
and $2.0 \leq d \leq 2.745$ has pairs of co-existing attractors,
the basins of which are separated by a boundary that may be either
smooth or fractal, depending on the choice of parameter values. We
choose for our studies $b\,=\,0.2$ and $d\,=\,2.7$, which
corresponds to there being two co-existing CAs separated by a
locally disconnected FBB (see Fig.~\ref{fig1}). The fractal
dimension of the boundary has been determined numerically (dim =
1.84472) by using the ``uncertainty exponent'' technique
introduced in \cite{Grebogi:85}. The chaotic attractors in
(\ref{hmap}) appear as the result of a period-doubling cascade
and, for the parameters chosen, each of them consists of two
disconnected parts.

\begin{figure}
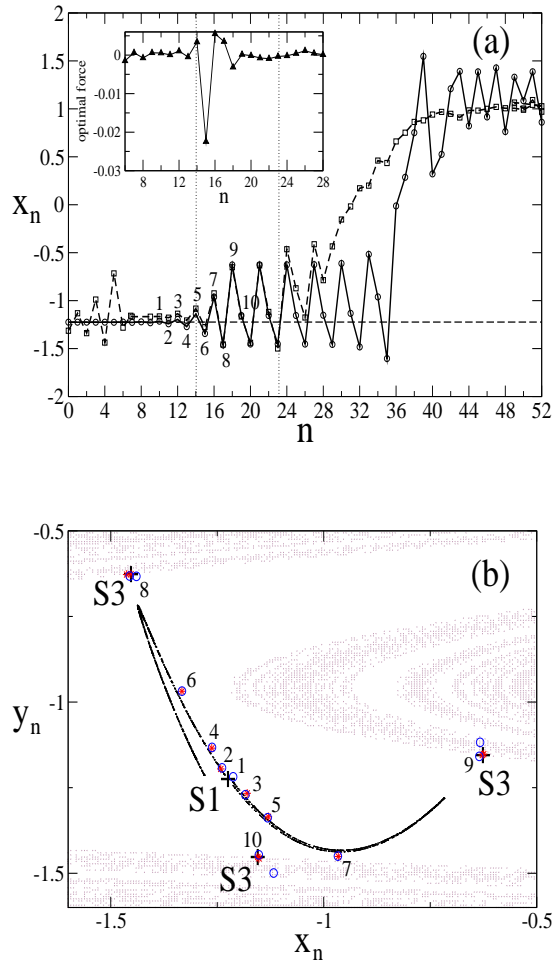

\includegraphics[width=7.2cm, height=5.85cm]{figure2a.eps}
\vskip 1cm
\includegraphics[width=7.2cm, height=5.85cm]{figure2b.eps}
\caption{\label{fig2} (a) the most probable escape path (dashed
line) connecting the CA with the period-3 saddle cycle lying on
the fractal boundary, obtained from the Monte-Carlo simulations
with $D\, =\, 10^{-5}.$ The optimal path found by the solution of
the boundary-value problem is shown as a solid line. The
$x$-coordinate of the saddle point S1 is shown by the horizontal
dashed line. (b) A two-dimensional plot of the paths presented in
(a) where the results obtained by solution of the the boundary
value problem (consecutively numbered points indicated by circles,
corresponding to the numbered points in (a)) coincide almost
perfectly with those obtained by numerical simulation (stars).
Inset in (a): the optimal force as determined in the numerical
simulation.}
\end{figure}

We have modelled the system (\ref{hmap}) numerically, exciting it
with weak noise, and have collected trajectories that include
escape paths from one CA to the other, and also the corresponding
realisations of noise that induced these transitions. By ensemble
averaging a few hundred such escape trajectories and noise
realisations, we have obtained the optimal escape path and the
corresponding optimal force, which are shown in Fig.~\ref{fig2}.
The results of this statistical analysis allow us both to
determine the boundary conditions near the CA and the FBB, and to
demonstrate the uniqueness of the MPEP. It can be seen in
particular that, in leaving the CA, the system (\ref{hmap}) falls
into a small neighbourhood of the saddle point of period 1 (S1)
located between its two disconnected parts and having the
multipliers $\rho_1=0.118975$ and $\rho_2=1.681025$. Its stable
manifolds separate the parts of the CA, while the unstable ones
belong to the CA. The system makes a few iterations in the
neighbourhood of S1 (initial plateau in Fig.~\ref{fig2}(a)) and
then moves to the FBB in three steps, crossing it at a saddle
point of period 3 (S3) with multipliers $\rho_1=0.001016$ and
$\rho_2=7.875512$. Calculations have shown that, for the chosen
parameter values, S3 lies within the FBB. Moreover, its stable
manifold (solid black line in Fig.~\ref{fig1}) is dense in the FBB
and detaches the open neighborhood, including an attractor, from
the FBB itself. This allows us to classify it as an accessible
boundary point~\cite{Grebogi:87}.

An analysis of the structure of escape paths inside the FBB has
shown that the homoclinic saddle points play a key role in its
formation. In the system (\ref{hmap}), we observe an infinite
sequence of saddle-node bifurcations of period $3,4,5,6,7...$,
which occur at parameter values $d_3<d_4<d_5<d_6<d_7...\,$ and are
caused by tangencies of the stable and unstable manifolds of the
saddle point O at the origin. The homoclinic orbits appearing as a
result of these bifurcations were classified earlier as {\em
original saddles}, and it was also shown that their stable and
unstable manifolds cross each other in a hierarchical
sequence~\cite{Grebogi:87}. To characterize this hierarchical
relation between original saddles, it is reasonable to introduce a
parameter $\mu$ equal to the ratio
$\mid\,\ln(\rho_1(P))\,\mid/\ln(\rho_2(P))$, where $\rho_1$ and $\rho_2$
are the multipliers of a saddle point $P$.
Calculations have shown that, for the original saddles of
period $3,4,5,6,7,8...$ in (\ref{hmap}), the following
hierarchical sequence of index $\mu$ values occurs:
$\mu_3=3.339,\,\mu_4=3.08,\,\mu_5=2.999,\,\mu_6=2.339,\,\mu_7=1.958,
\,\mu_8=1.5 39$. Moreover, the values of index $\mu$ corresponding
to the other homoclinic saddle cycles are close to zero.
Correspondingly the probability of finding the system in their
neighbourhood tends to zero.

These results allows us to infer features of a fluctuational
transition through a locally disconnected FBB that are probably
generic, as follows: (i) it always occurs through a unique
accessible boundary point; and (ii) the original saddles forming
the homoclinic structure of the system play a key role in the
formation of the paths inside the FBB, the difference in their
local stability defining the hierarchical relationship between
them. Thus, we may claim that complicated structure of escape
trajectories, caused by the thin homoclinic structure and their
randomness, has in many respects a deterministic nature.

Having now understood the mechanism of escape, we can seek the
MPEP. According to the Hamiltonian theory of
fluctuations~\cite{Kautz:87,Beale:89,Grassberger:89,Hamm:91} the
MPEP is the path which minimizes the energy
\begin{equation}
S=\frac{1}{2}\sum_{n=1}^{N}\: {\xi^{T}_n\, \xi_n},
\label{energy}
\end{equation}
of the possible realizations of noise $\{\xi_n\}$ inducing a
transition of the system (\ref{hmap}) from the CA (with the
initial condition on S1) to the FBB (with the final condition on
the accessible orbit S3). The Lagrangian of the corresponding
variational problem can be found following~\cite{Grassberger:89}
(cf.\ \cite{Dykman:90}) in the form
\[
L=\frac{1}{2}\sum_{n=1}^{N}\, {\bf{\xi^{T}_n}\, \xi_n} +
\sum_{n=1}^{N}\, {\bf \lambda_n^T\, (x_{n+1}-f(x_n)-\xi_n)},
\]
where (\ref{hmap}) is taken into account using the Lagrange
multiplier $\lambda_n$. Varying $L$ with respect to $\xi_n$,
$\lambda_n$, and $x_n$, the following area-preserving map is
obtained:
\begin{eqnarray}
\label{ext_holmes} \nonumber
x_{n+1}&=& y_{n}\\
y_{n+1}&=& -b\, x_{n} + d\, y_{n}-y_{n}^{3}+\lambda_n^{y}\\
\nonumber \lambda_{n+1}^{x}&=&(d-3x_{n+1}^{2})\, \lambda_{n}^{x}/b
- \lambda_{n}^{y}/b\\ \nonumber
\lambda_{n+1}^{y}&=&\lambda_{n}^{x}\\ \nonumber
\end{eqnarray}
Equations (\ref{ext_holmes}) are supplemented by the following
boundary conditions
\begin{equation}
\lim_{n\rightarrow -\infty} \lambda_n^y=0,\: \;
(x_n^0,y_n^0) \in S1 , \: \; (x_n^1,y_n^1) \in S3.
\end{equation}
The MPEP is the minimum-energy heteroclinic trajectory connecting
S1 to S3 in the phase space of (\ref{ext_holmes}). The solution of
this boundary value problem is in general complicated, because of
the presence of multiple local energy minima \cite{Luchinsky:02a}
induced by the complex geometrical structure of the unstable
manifolds of S3 (see e.g.~\cite{Graham:84a} for a discussion). The
solution of the boundary value problem involves a parameterization
of the structure of the multiple local minima requiring, in turn,
a proper parameterization of the unstable manifold in the vicinity
of the initial conditions \cite{Beri:03}. The MPEP found by this
method is shown in Fig.~\ref{fig2}. It can be seen that, within
the range shown by the vertical dotted lines in
Fig.~\ref{fig2}(a), the theoretical MPEP closely coincides with
the path obtained by statistical analysis of escape trajectories
in the Monte Carlo simulations.
Note that no further action is required to bring the system to the
other attractor once it has reached the accessible orbit of the
FBB, i.e.\ once it has reached the points numbered 8 in
Fig.~\ref{fig2}(a) and (b); correspondingly, the optimal force
measured in the numerical simulations (inset) falls back to zero.

\begin{figure}
\includegraphics[width=7.1cm, height=6cm]{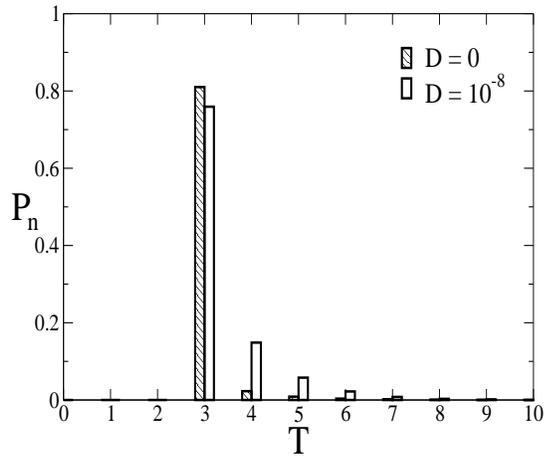}
\caption{\label{fig3} Probabilities of finding a fragment
corresponding to the different period-$T$ original saddle cycle in
the collected escape trajectories.}
\end{figure}

The existence of an almost deterministic mechanism of transition
across the FBB raises important questions about the effect of
noise on this mechanism, and on the structure of escape
trajectories inside the FBB. We have therefore used randomly
chosen initial conditions in a very small neighborhood of the
accessible point S3 through which escape occurs (see Fig.~2(b)).
By definition, any arbitrarily small neighborhood of S3, lies
within the FBB, and must contain points belonging to the basins of
both attractors. Therefore the system can cross the FBB starting
from a very small neighborhood of S3, even in the absence of
noise. By collecting all such successful escape paths, we have
calculated the probabilities for the system to pass via small
neighborhoods of different original saddle cycles during its
escape, both in the presence and absence of noise. As can be seen
from Fig.~\ref{fig3}, the corresponding probabilities demonstrate
the same hierarchical interrelationship in both cases, which is
determined by the value of index $\mu$ defined above. This
structure is evidently robust with respect to noise-induced
perturbations. The addition of noise causes a slight broadening of
the distribution in Fig.~~\ref{fig3} and a small increase of the
probability for the system to escape via original saddles of
larger period.

In conclusion, we have described the mechanism by which
noise-induced escape occurs through a locally disconnected FBB. We
have found the (unique) most probable escape path from  a chaotic
attractor to the fractal boundary, using both statistical analyses
of fluctuational trajectories, and the Hamiltonian theory of
fluctuations. We have shown that the original saddles forming the
homoclinic structure play a key role in effecting the transition
through the FBB itself. In particular, their local stability
defines the hierarchical relationship between the probabilities
for the system to pass via small neighborhoods of different
original saddle cycles during its escape, both in the presence and
in the absence of noise. We emphasize that the escape mechanism we
have revealed must be applicable to the broad class of two
dimensional maps and flows
\cite{Holmes:79,Cartwright:51,Yorke:85a} that exhibit the same
type of FBB. For instance, one possible application of our results
is to the development of an energy-optimal control scheme for the
CO$_2$ laser, a discrete model of which demonstrates the type of
FBB considered above~\cite{Grigorieva}.

The authors would like to thank Ulrike Feudel, Igor Khovanov and
Suso Kraut for fruitful discussions. The work was supported by the
Engineering and Physical Sciences Research Council (UK), the
Russian Foundation for Fundamental Science, and INTAS.

\end{document}